\documentstyle[aps,prd]{revtex}
\begin{document}
\setlength{\topmargin}{-2cm}
\setlength{\oddsidemargin}{-1cm}
\setlength{\evensidemargin}{-1cm}
\newcommand{\nc}{\newcommand}
\nc{\beq}{\begin{equation}}
\nc{\eeq}{\end{equation}}
\nc{\bea}{\begin{eqnarray}}
\nc{\eea}{\end{eqnarray}}
\nc{\ba}{\begin{array}}
\nc{\ea}{\end{array}}
\nc{\nn}{\nonumber}
\nc{\bpi}{\begin{picture}}
\nc{\epi}{\end{picture}}
\nc{\scs}{\scriptstyle}
\nc{\sss}{\scriptscriptstyle}
\nc{\sst}{\scriptstyle}
\nc{\ts}{\textstyle}
\nc{\ds}{\displaystyle}
\nc{\lra}{\leftrightarrow}
\nc{\lan}{\langle}
\nc{\ran}{\rangle}
\nc{\Int}[1]{{\ds\int_{#1}}}
\nc{\msbar}{$\overline{\mb{MS}}$}

\nc{\al}{\alpha}
\nc{\be}{\beta}
\nc{\ga}{\gamma}
\nc{\Ga}{\Gamma}
\nc{\de}{\delta}
\nc{\De}{\Delta}
\nc{\ep}{\epsilon}
\nc{\ve}{\varepsilon}
\nc{\eb}{\bar{\eta}}
\nc{\et}{\eta}
\nc{\ka}{\kappa}
\nc{\la}{\lambda}
\nc{\La}{\Lambda}
\nc{\vt}{\vartheta}
\nc{\rh}{\rho}
\nc{\si}{\sigma}
\nc{\Si}{\Sigma}
\nc{\th}{\theta}
\nc{\Th}{\Theta}
\nc{\ze}{\zeta}
\nc{\p}{\partial}

\nc{\mb}[1]{\mbox{#1}}
\nc{\Tr}{\mb{Tr}}
\nc{\Wb}{\mb{\boldmath $\ds W$}}
\nc{\Vb}{\mb{\boldmath $\ds V$}}
\nc{\mw}{m_{\sss W0}}
\nc{\mz}{m_{\sss Z0}}
\nc{\half}{{\ts\frac{1}{2}}}
\nc{\ihalf}{{\ts\frac{i}{2}}}
\nc{\quarter}{{\ts\frac{1}{4}}}
\nc{\dg}{\dagger}
\nc{\Lag}{{\cal L}}
\nc{\od}{{\cal O}}
\nc{\mubar}{\bar{\mu}}
\nc{\kba}{\bar{k}}
\def\ltap{\;\raisebox{-.4ex}{\rlap{$\sim$}}\raisebox{.4ex}{$<$}\;}
\def\gtap{\;\raisebox{-.4ex}{\rlap{$\sim$}}\raisebox{.4ex}{$>$}\;}

\begin{flushright}
\normalsize
Freiburg-THEP 98/19\\
hep-ph/9809520\\
September 1998\vspace{0.8cm}\\
\end{flushright}

\begin{center}
{\Large\bf Resonance in Strong \boldmath$WW$ Rescattering\vspace{0.2cm}\\
in Massive SU(2) Gauge Theory}
\vspace{1.2cm}\\
J.J.\ van der Bij and Boris Kastening\vspace{0.4cm}\\
\it Albert-Ludwigs-Universit\"at Freiburg\\
\it Fakult\"at f\"ur Physik\\
\it Hermann-Herder-Stra\ss e 3\\
\it D-79104 Freiburg\\
\it Germany\vspace{1.2cm}\\
\end{center}

\begin{abstract}
We investigate the effects of $WW$ rescattering through strong anomalous
four-vector boson couplings.
In the $I=1$, $J=1$ channel, we find a resonance with a mass of approximately
$200\mb{GeV}$ and a width of less than $12\mb{GeV}$.
In an application to pion physics we find a small correction to the KSRF
relation.
\end{abstract}

\section{Introduction}
Within the standard model of electroweak interactions the gauge principle
fully determines the selfcouplings of the vector bosons.
Therefore the measurement of these couplings is of prime importance,
as deviations would be an indication of new physics.
The gauge principle predicts well-defined three and four-vector boson
selfcouplings.
Deviations of these values can be described in a gauge-invariant way in
the St\"uckelberg formalism \cite{stuc}.
Within the St\"uckelberg formalism the standard model is described as a
gauged nonlinear sigma model.
This implicitly assumes that the Higgs particle does not play a fundamental
role.
Alternatively one can see it as the $m_H\rightarrow\infty$ limit of the
standard model.
If indeed anomalous vector boson couplings are present, this should be a
reasonable assumption, since in that case strong interactions should be
present.
The triple vector boson couplings are severely constrained by the
measurements at the CERN $e^+e^-$ collider LEP-200 and the Fermilab
Tevatron \cite{part}.
Also, indirect limits from LEP-100 and $(g-2)_{\mu}$ exist.
Altogether experiments indicate they should be small.
This is not too surprising, as it is very hard to construct a model that
would give rise to large effects.
The reason is that within the three-vector boson couplings there is always
an interplay between longitudinal and transversal vector bosons.

For the four-vector boson couplings the situation is somewhat different.
Here one can write down vertices that contain longitudinal vector bosons
only.
These vertices correspond to the Goldstone boson sector of the theory
and it is much easier to generate strong interactions in this sector.
Such results come typically through intermediate heavy Higgs boson exchange.
An example of such a model is given in \cite{hillvelt}, where the
strong interactions are generated via singlet effects in the Higgs sector.
Also, in the standard model the two-loop heavy Higgs correction in the
four-vector boson \cite{jikia} couplings is an order of magnitude larger
than in the three-vector boson couplings \cite{jochum}.
About the four-vector boson couplings much less is known than about the
three-vector boson couplings.
Direct experiments probing these interactions do not exist at present.
They can at the moment only be tested through radiative corrections in the
$\rho$ parameter.
These corrections can be calculated within perturbation theory with a cutoff
$\La$ .
Within the four-vector boson couplings one should distinguish between two
types.
In the standard model there is an extra global SU$_R$(2) symmetry
when the hypercharge is turned off.
For the anomalous couplings this is not necessarily the case.
The couplings that violate SU$_R$(2) even in the
absence of hypercharge give quartically divergent corrections to
$\de\rho$ \cite{quart,kast} and should therefore be negligibly
small.
Physically, this means that the underlying strong interactions, generating the
anomalous couplings, should respect the SU$_R$(2) invariance.
This leaves only two operators that preserve SU$_R$(2) invariance in the
absence of hypercharge.
If one could use simple cutoff perturbation theory these operators still
give quadratically divergent corrections to $\de\rho$ \cite{quart,kast}
and are sufficiently suppressed to give only small effects in future
colliders.
In \cite{dawson} the quadratic divergences were ignored and therefore the
limits are weak.

This leaves only the possibility that the anomalous couplings are so large
that perturbation theory cannot be trusted and therefore the low energy
limits are invalidated.
It is precisely this case that we study in this paper.
We assume that anomalous couplings are present which preserve the SU$_R$(2)
symmetry of the theory and assume the coefficients for the corresponding
operators to be large.
To simplify the calculation we actually ignore hypercharge altogether and
work in the SU$_L$(2)$\times$SU$_R$(2) model.
This leads to a simplification because we can study different isospin
channels separately.
Because the interactions are assumed to be strong, vector boson scattering
cannot be described by the tree-level vertex.
We therefore perform a resummation of loop graphs.
In most channels we find no particularly interesting effect.
The only exception is the $I=1$, $J=1$ channel, where a resonance is found.
In the case of very strong anomalous couplings, which we assume, the
resonance can be quite close to the two-vector boson threshold.
The coupling of the resonance
to the vector bosons is found to be suppressed by the cutoff and could be
small.
Dependent on the parameters the resonance could be visible even at LEP-200.

The paper is organized as follows.
In section \ref{model} we present the model.
In section \ref{bsum} we perform the calculation of the bubble sum.
In section \ref{discussion} we discuss the results.
The appendices contain technical details.

\section{The Model}
\label{model}

\subsection{Lagrangian}
We work in a pure massive SU(2) gauge theory and introduce the anomalous
couplings in a gauge-invariant way using the St\"uckelberg formalism
\cite{stuc}.
That is, we write the theory as a gauged nonlinear sigma model:
\beq
\Lag_{gauge}=-\half\Tr(\Wb_{\mu\nu}\Wb^{\mu\nu})+\mw^2\Tr(\Vb_\mu\Vb^\mu)
\eeq
with
\beq
\Wb_\mu\equiv\half\tau_aW_\mu^a=\Wb_\mu^\dg\,,
\eeq
\beq
\Wb_{\mu\nu}\equiv\half\tau_aW_{\mu\nu}^a
=\p_\mu\Wb_\nu-\p_\nu\Wb_\mu+ig[\Wb_\mu,\Wb_\nu]=\Wb_{\mu\nu}^\dg\,,
\eeq
\beq
\ts\Vb_\mu\equiv-\frac{i}{g}(D_\mu U)U^\dg=\Vb_\mu^\dg\,,
\eeq
\beq
D_\mu U\equiv\p_\mu U+ig\Wb_\mu U\,,
\eeq
\beq
U=\exp(iv_a\tau_a)
\eeq
with real fields $v_a$, i.e., the SU(2)-valued field $U$ describes the
Goldstone degrees of freedom.
$\mw$ is the mass of the three gauge bosons in absence of the higher
covariant derivative terms to be added below.
The anomalous couplings are then introduced as
\bea
\Lag_{\mb{\scriptsize\em ano}}
&=&
g_4(\Tr[\Vb_\mu\Vb_\nu])^2+g_5(\Tr[\Vb^\mu\Vb_\mu])^2
\nn\\
&=&
g_4(\Tr[\Vb_\mu\Vb_\nu^\dg])^2+g_5(\Tr[\Vb^\mu\Vb_\mu^\dg])^2
\nn\\
&=&\ts
g_4g^{-4}\{\Tr[(D_\mu U)(D_\nu U)^\dg]\}^2
+g_5g^{-4}\{\Tr[(D^\mu U)(D_\mu U)^\dg]\}^2\,.
\eea
Their form is determined by the requirement that they conserve {\it CP},
are not accompanied by three-vector boson couplings (on which the
limits are much more stringent, as was mentioned in the introduction)
and that they are invariant under the custodial SU$_R$(2) symmetry
$U\rightarrow UU_R$ with $U\in SU(2)$.

To regulate higher-than-logarithmic divergences, we introduce higher
covariant derivative terms through
\beq
\label{hicoder}
\Lag_{hcd}=\frac{1}{2\La_W^2}\Tr[(D_\al\Wb_{\mu\nu})(D^\al\Wb^{\mu\nu})]
-\frac{\mw^2}{\La_V^2}\Tr[(D_\al\Vb_\mu)(D^\al\Vb^\mu)]
\eeq
with
\beq
D_\al\Wb_{\mu\nu}=\p_\al\Wb_{\mu\nu}+ig[\Wb_\al,\Wb_{\mu\nu}]\,,
\eeq
$\La_W$ and $\La_V$ effectively being momentum-cutoffs. These are the
unique dimension 6 higher derivative propagator terms and are 
further discussed in \cite{kast}.

We work in unitary gauge, where we have $U=\mb{\bf 1}$, $v_a=0$,
and therefore
\beq
\Vb_\mu=\ts\Wb_\mu\,.
\eeq

The signature of our metric is
\beq
g_{\mu\nu}=\mb{diag}(1,-1,-1,-1)\,.
\eeq

\subsection{Feynman Rules}
The $W$ propagator in unitary gauge with higher covariant derivatives is
\beq\ts
\De_{\mu\nu}^W(k)
=\De_{tr}^W(k^2)\left(g_{\mu\nu}-\frac{k_\mu k_\nu}{k^2}\right)
+\De_{lg}^W(k^2)\frac{k_\mu k_\nu}{k^2}
\eeq
with
\beq
\De_{tr}^W(k^2)=\frac{i\La_W^2}{(k^2-m_-^2)(k^2-m_+^2)}\,,
\eeq
\beq
\De_{lg}^W(k^2)=-\frac{i\La_V^2}{\mw^2}\frac{1}{k^2-\La_V^2}\,,
\eeq
\beq
m_\pm^2
=\frac{\La_W^2}{2}\left[\left(1+\frac{\mw^2}{\La_V^2}\right)
\pm\sqrt{\left(1+\frac{\mw^2}{\La_V^2}\right)^2-\frac{4\mw^2}{\La_W^2}}
\right]
=\left\{\ba{l}\La_W^2+\od(\La^0)\\\mw^2+\od(\La^{-2})\ea\right.\,.
\eeq

The Feynman rule for the anomalous four-vector boson couplings is
\bea
\label{vertex}
\rule[0pt]{0pt}{25pt}
\bpi(75,0)(-3,17)
\multiput(36,21)(4,4){4}{\oval(4,4)[br]}
\multiput(40,21)(4,4){4}{\oval(4,4)[tl]}
\multiput(20,33)(4,-4){4}{\oval(4,4)[tr]}
\multiput(24,33)(4,-4){4}{\oval(4,4)[bl]}
\multiput(20,5)(4,4){4}{\oval(4,4)[br]}
\multiput(24,5)(4,4){4}{\oval(4,4)[tl]}
\multiput(36,17)(4,-4){4}{\oval(4,4)[tr]}
\multiput(40,17)(4,-4){4}{\oval(4,4)[bl]}
\put(36,19){\circle*{3}}
\put(15,35){\makebox(0,0)[r]{$a,\al$}}
\put(15,5){\makebox(0,0)[r]{$b,\be$}}
\put(55,35){\makebox(0,0)[l]{$c,\ga$}}
\put(55,5){\makebox(0,0)[l]{$d,\de$}}
\epi
&=&\ts
i\{\;\,\de_{ab}\de_{cd}[2g_5g_{\al\be}g_{\ga\de}
+g_4(g_{\al\ga}g_{\be\de}+g_{\al\de}g_{\be\ga})]
\nn\\
&&\ts
\;\;+\de_{ac}\de_{bd}[2g_5g_{\al\ga}g_{\be\de}
+g_4(g_{\al\be}g_{\ga\de}+g_{\al\de}g_{\be\ga})]
\nn\\
&&\ts
\;\;+\de_{ad}\de_{bc}[2g_5g_{\al\de}g_{\be\ga}
+g_4(g_{\al\ga}g_{\be\de}+g_{\al\be}g_{\ga\de})]\}
\nn\\
&\equiv&
V_{\al\be\ga\de}^{abcd}\,.
\eea

\section{Bubble Sum}
\label{bsum}

\subsection{Definitions}
Assuming the anomalous couplings to dominate the gauge coupling, we
compute the ``bubble sum''
\bea
\label{bubblesum}
\rule[-25pt]{0pt}{45pt}
\bpi(103,0)(-3,17)
\put(15,35){\makebox(0,0)[r]{$a,\al$}}
\put(15,5){\makebox(0,0)[r]{$b,\be$}}
\put(20,5){\line(1,1){15}}
\put(20,35){\line(1,-1){15}}
\put(35,20){\circle*{3}}
\put(49,20){\circle{28}}
\put(49,20){\makebox(0,0){$\Sigma$}}
\put(63,20){\circle*{3}}
\put(63,20){\line(1,1){15}}
\put(63,20){\line(1,-1){15}}
\put(83,35){\makebox(0,0)[l]{$c,\ga$}}
\put(83,5){\makebox(0,0)[l]{$d,\de$}}
\epi
&\equiv&
\rule[-25pt]{0pt}{45pt}
\bpi(75,0)(-3,17)
\put(15,35){\makebox(0,0)[r]{$a,\al$}}
\put(15,5){\makebox(0,0)[r]{$b,\be$}}
\put(20,5){\line(1,1){15}}
\put(20,35){\line(1,-1){15}}
\put(35,20){\circle*{3}}
\put(35,20){\line(1,1){15}}
\put(35,20){\line(1,-1){15}}
\put(55,35){\makebox(0,0)[l]{$c,\ga$}}
\put(55,5){\makebox(0,0)[l]{$d,\de$}}
\epi
+
\bpi(103,0)(-3,17)
\put(15,35){\makebox(0,0)[r]{$a,\al$}}
\put(15,5){\makebox(0,0)[r]{$b,\be$}}
\put(20,5){\line(1,1){15}}
\put(20,35){\line(1,-1){15}}
\put(35,20){\circle*{3}}
\put(49,20){\circle{28}}
\put(63,20){\circle*{3}}
\put(63,20){\line(1,1){15}}
\put(63,20){\line(1,-1){15}}
\put(83,35){\makebox(0,0)[l]{$c,\ga$}}
\put(83,5){\makebox(0,0)[l]{$d,\de$}}
\epi
+
\bpi(131,0)(-3,17)
\put(15,35){\makebox(0,0)[r]{$a,\al$}}
\put(15,5){\makebox(0,0)[r]{$b,\be$}}
\put(20,5){\line(1,1){15}}
\put(20,35){\line(1,-1){15}}
\put(35,20){\circle*{3}}
\put(49,20){\circle{28}}
\put(63,20){\circle*{3}}
\put(77,20){\circle{28}}
\put(91,20){\circle*{3}}
\put(91,20){\line(1,1){15}}
\put(91,20){\line(1,-1){15}}
\put(111,35){\makebox(0,0)[l]{$c,\ga$}}
\put(111,5){\makebox(0,0)[l]{$d,\de$}}
\epi
+
\cdots
\nn\\
&\equiv&
\rule[-25pt]{0pt}{45pt}
\bpi(103,0)(-3,17)
\put(15,35){\makebox(0,0)[r]{$a,\al$}}
\put(15,5){\makebox(0,0)[r]{$b,\be$}}
\put(20,5){\line(1,1){15}}
\put(20,35){\line(1,-1){15}}
\put(35,20){\circle*{3}}
\put(49,20){\circle{28}}
\put(49,20){\makebox(0,0){0}}
\put(63,20){\circle*{3}}
\put(63,20){\line(1,1){15}}
\put(63,20){\line(1,-1){15}}
\put(83,35){\makebox(0,0)[l]{$c,\ga$}}
\put(83,5){\makebox(0,0)[l]{$d,\de$}}
\epi
+
\bpi(103,0)(-3,17)
\put(15,35){\makebox(0,0)[r]{$a,\al$}}
\put(15,5){\makebox(0,0)[r]{$b,\be$}}
\put(20,5){\line(1,1){15}}
\put(20,35){\line(1,-1){15}}
\put(35,20){\circle*{3}}
\put(49,20){\circle{28}}
\put(49,20){\makebox(0,0){1}}
\put(63,20){\circle*{3}}
\put(63,20){\line(1,1){15}}
\put(63,20){\line(1,-1){15}}
\put(83,35){\makebox(0,0)[l]{$c,\ga$}}
\put(83,5){\makebox(0,0)[l]{$d,\de$}}
\epi
+
\bpi(103,0)(-3,17)
\put(15,35){\makebox(0,0)[r]{$a,\al$}}
\put(15,5){\makebox(0,0)[r]{$b,\be$}}
\put(20,5){\line(1,1){15}}
\put(20,35){\line(1,-1){15}}
\put(35,20){\circle*{3}}
\put(49,20){\circle{28}}
\put(49,20){\makebox(0,0){2}}
\put(63,20){\circle*{3}}
\put(63,20){\line(1,1){15}}
\put(63,20){\line(1,-1){15}}
\put(83,35){\makebox(0,0)[l]{$c,\ga$}}
\put(83,5){\makebox(0,0)[l]{$d,\de$}}
\epi
+\cdots
\nn\\
&\equiv&
\,B_{\al\be\ga\de}^{(0)abcd}
+\,B_{\al\be\ga\de}^{(1)abcd}
+\,B_{\al\be\ga\de}^{(2)abcd}
+\cdots
\nn\\
&\equiv&
\Sigma_{\al\be\ga\de}^{abcd}
\,,
\eea
where all lines represent gauge bosons and where the vertices contain
only the anomalous part.
Note that (\ref{bubblesum}) is invariant under each of
\bea
\label{sym1}
(a,\al)&\lra&(b,\be)\,,
\\
\label{sym2}
(c,\ga)&\lra&(d,\de)\,,
\\
\label{sym3}
(a,\al,b,\be)&\lra&(c,\ga,d,\de)\,.
\eea

To express equation (\ref{bubblesum}) in a purely algebraic form, we
split each additional bubble into a vertex part (\ref{vertex}) and a
propagator part
\beq
\bpi(75,0)(-3,17)
\put(15,35){\makebox(0,0)[r]{$a,\al$}}
\put(15,5){\makebox(0,0)[r]{$b,\be$}}
\put(20,5){\line(1,0){7.5}}
\put(20,35){\line(1,0){7.5}}
\put(27.5,0){\framebox(15,40){$\De$}}
\put(50,5){\line(-1,0){7.5}}
\put(50,35){\line(-1,0){7.5}}
\put(55,35){\makebox(0,0)[l]{$c,\ga$}}
\put(55,5){\makebox(0,0)[l]{$d,\de$}}
\epi
\equiv
\frac{1}{2}
\left(
\rule[-25pt]{0pt}{45pt}
\bpi(75,0)(-3,17)
\put(15,35){\makebox(0,0)[r]{$a,\al$}}
\put(15,5){\makebox(0,0)[r]{$b,\be$}}
\put(20,5){\line(1,0){30}}
\put(20,35){\line(1,0){30}}
\put(55,35){\makebox(0,0)[l]{$c,\ga$}}
\put(55,5){\makebox(0,0)[l]{$d,\de$}}
\epi
+
\bpi(75,0)(-3,17)
\put(15,35){\makebox(0,0)[r]{$a,\al$}}
\put(15,5){\makebox(0,0)[r]{$b,\be$}}
\put(20,5){\line(1,1){10.5}}
\put(39.5,24.5){\line(1,1){10.5}}
\put(39.5,15.5){\oval(18,18)[tl]}
\put(20,35){\line(1,-1){30}}
\put(55,35){\makebox(0,0)[l]{$c,\ga$}}
\put(55,5){\makebox(0,0)[l]{$d,\de$}}
\epi
\right)\,,
\eeq
where we have imposed the relevant symmetries (\ref{sym1})-(\ref{sym3}).
equation (\ref{bubblesum}) can now be written as
\bea
\lefteqn{\rule[-25pt]{0pt}{45pt}
\bpi(103,0)(-3,17)
\put(15,35){\makebox(0,0)[r]{$a,\al$}}
\put(15,5){\makebox(0,0)[r]{$b,\be$}}
\put(20,5){\line(1,1){15}}
\put(20,35){\line(1,-1){15}}
\put(35,20){\circle*{3}}
\put(49,20){\circle{28}}
\put(49,20){\makebox(0,0){$\Sigma$}}
\put(63,20){\circle*{3}}
\put(63,20){\line(1,1){15}}
\put(63,20){\line(1,-1){15}}
\put(83,35){\makebox(0,0)[l]{$c,\ga$}}
\put(83,5){\makebox(0,0)[l]{$d,\de$}}
\epi}
\nn\\
&=&
\rule[-25pt]{0pt}{45pt}
\bpi(75,0)(-3,17)
\put(15,35){\makebox(0,0)[r]{$a,\al$}}
\put(15,5){\makebox(0,0)[r]{$b,\be$}}
\put(20,5){\line(1,1){15}}
\put(20,35){\line(1,-1){15}}
\put(35,20){\circle*{3}}
\put(35,20){\line(1,1){15}}
\put(35,20){\line(1,-1){15}}
\put(55,35){\makebox(0,0)[l]{$c,\ga$}}
\put(55,5){\makebox(0,0)[l]{$d,\de$}}
\epi
+
\bpi(118,0)(-3,17)
\put(15,35){\makebox(0,0)[r]{$a,\al$}}
\put(15,5){\makebox(0,0)[r]{$b,\be$}}
\put(20,5){\line(1,1){15}}
\put(20,35){\line(1,-1){15}}
\put(35,20){\circle*{3}}
\put(49,20){\oval(28,28)[l]}
\put(49,0){\framebox(15,40){$\De$}}
\put(64,20){\oval(28,28)[r]}
\put(78,20){\circle*{3}}
\put(78,20){\line(1,1){15}}
\put(78,20){\line(1,-1){15}}
\put(98,35){\makebox(0,0)[l]{$c,\ga$}}
\put(98,5){\makebox(0,0)[l]{$d,\de$}}
\epi
+
\bpi(161,0)(-3,17)
\put(15,35){\makebox(0,0)[r]{$a,\al$}}
\put(15,5){\makebox(0,0)[r]{$b,\be$}}
\put(20,5){\line(1,1){15}}
\put(20,35){\line(1,-1){15}}
\put(35,20){\circle*{3}}
\put(49,20){\oval(28,28)[l]}
\put(49,0){\framebox(15,40){$\De$}}
\put(64,20){\oval(28,28)[r]}
\put(78,20){\circle*{3}}
\put(92,20){\oval(28,28)[l]}
\put(92,0){\framebox(15,40){$\De$}}
\put(107,20){\oval(28,28)[r]}
\put(121,20){\circle*{3}}
\put(121,20){\line(1,1){15}}
\put(121,20){\line(1,-1){15}}
\put(141,35){\makebox(0,0)[l]{$c,\ga$}}
\put(141,5){\makebox(0,0)[l]{$d,\de$}}
\epi
+\cdots\,.
\eea
Since the vertex is independent of any momenta, the integration over the
loop momenta can be done in the propagator part.
Abbreviate
\beq
\label{int}
\Int{p}\equiv\mu^{4-d}\int\frac{d^dp} {(2\pi)^d}
\eeq
with the renormalization scale $\mu$ and use dimensional regularization
together with the modified minimal subtraction (\msbar) scheme throughout.
Define
\bea
\label{ptensor}
P_{\al\be\ga\de}^{abcd}
&\equiv&
\half\Int{p}
\;\;
\rule[-25pt]{0pt}{45pt}
\bpi(75,0)(-3,17)
\put(15,35){\makebox(0,0)[r]{$a,\al$}}
\put(15,5){\makebox(0,0)[r]{$b,\be$}}
\put(20,5){\line(1,0){7.5}}
\put(20,35){\line(1,0){7.5}}
\put(27.5,0){\framebox(15,40){$\De$}}
\put(50,5){\line(-1,0){7.5}}
\put(50,35){\line(-1,0){7.5}}
\put(55,35){\makebox(0,0)[l]{$c,\ga$}}
\put(55,5){\makebox(0,0)[l]{$d,\de$}}
\epi
\nn\\
&=&\ts
\frac{1}{4}\Int{p}\Big\{\hspace{5pt}\de_{ac}\de_{bd}
\left[\De_{tr}^W(p^2)\left(g_{\al\ga}-\frac{p_\al p_\ga}{p^2}\right)+
\De_{lg}^W(p^2)\frac{p_\al p_\ga}{p^2}\right]
\nn\\
&&\hspace{30pt}\ts
\times\left[\De_{tr}^W((p+k)^2)\left(g_{\be\de}
-\frac{(p+k)_\be (p+k)_\de}{(p+k)^2}\right)
+\De_{lg}^W((p+k)^2)\frac{(p+k)_\be (p+k)_\de}{(p+k)^2}\right]
\nn\\
&&\hspace{20pt}\ts
+\de_{ad}\de_{bc}
\left[\De_{tr}^W(p^2)\left(g_{\al\de}-\frac{p_\al p_\de}{p^2}\right)
+\De_{lg}^W(p^2)\frac{p_\al p_\de}{p^2}\right]
\nn\\
&&\hspace{30pt}\ts
\times\left[\De_{tr}^W((p+k)^2)\left(g_{\be\ga}-
\frac{(p+k)_\be (p+k)_\ga}{(p+k)^2}\right)
+\De_{lg}^W((p+k)^2)\frac{(p+k)_\be (p+k)_\ga}{(p+k)^2}\right]
\Big\}
\nn\\
&\equiv&
\Big\{\hspace{10pt}
\de_{ac}\de_{bd}\Big[\De_1(k^2)g_{\al\ga}g_{\be\de}
+\De_2(k^2)(g_{\al\be}g_{\ga\de}+g_{\al\de}g_{\be\ga})
+\De_3(k^2)(g_{\al\ga}k_\be k_\de+g_{\be\de}k_\al k_\ga)/k^2
\nn\\
&&\hspace{40pt}
+\De_4(k^2)(g_{\al\be}k_\ga k_\de+g_{\ga\de}k_\al k_\be
+g_{\al\de}k_\be k_\ga+g_{\be\ga}k_\al k_\de)/k^2
+\De_5(k^2)k_\al k_\be k_\ga k_\de/k^4\Big]
\nn\\
&&\;\;
+\de_{ad}\de_{bc}\Big[\De_1(k^2)g_{\al\de}g_{\be\ga}
+\De_2(k^2)(g_{\al\be}g_{\ga\de}+g_{\al\ga}g_{\be\de})
+\De_3(k^2)(g_{\al\de}k_\be k_\ga+g_{\be\ga}k_\al k_\de)/k^2
\nn\\
&&\hspace{40pt}
+\De_4(k^2)(g_{\al\be}k_\ga k_\de+g_{\ga\de}k_\al k_\be
+g_{\al\ga}k_\be k_\de+g_{\be\de}k_\al k_\ga)/k^2
+\De_5(k^2)k_\al k_\be k_\ga k_\de/k^4\Big]
\Big\}\,,
\nn\\
\eea
where an additional factor of $1/2$ has been introduced to account for
the implicit symmetry factors in the bubble sum (\ref{bubblesum}), which
can now be written as
\beq
\label{vpsum}
\Sigma_{\al\be\ga\de}^{abcd}
=
V_{\al\be\ga\de}^{abcd}
+
V_{\al\be\mu\nu}^{abmn}
P_{\mu\nu\mu'\nu'}^{mnm'n'}
V_{\mu'\nu'\ga\de}^{m'n'cd}
+
V_{\al\be\mu\nu}^{abmn}
P_{\mu\nu\mu'\nu'}^{mnm'n'}
V_{\mu'\nu'\mu''\nu''}^{m'n'm''n''}
P_{\mu''\nu''\mu'''\nu'''}^{m''n''m'''n'''}
V_{\mu'''\nu'''\ga\de}^{m'''n'''cd}
+
\cdots\,.
\eeq
The strategy for computing $\De_1$ through $\De_5$ is given in the appendix.
Keeping only at least quadratically divergent, i.e.\ $\od(\La^2)$ terms
for the total $\De_1$ through $\De_5$, and $\od(\La^0)$
terms for their real parts, the results are
\bea
\label{de1}
\De_1(k^2)
&=&
-\frac{i\La_V^4}{96(4\pi)^2\mw^4}
\left[\frac{1}{\ep}+\frac{5}{6}-\ln\frac{\La_V^2}{\mubar^2}\right]
\nn\\
&&
-\frac{5i\La_W^2\La_V^2}{48(4\pi)^2(\La_W^2-\La_V^2)\mw^2}
\ln\frac{\La_W^2}{\La_V^2}
-\frac{7i\La_V^2}{1152(4\pi)^2\mw^2}\left(\frac{k^2}{\mw^2}\right)
+\od(\La^0)\,,
\eea

\beq
\mb{Re}\De_1(k^2)
=\frac{\pi\sqrt{1-4\mw^2/k^2}}{60(4\pi)^2}\left[
\left(\frac{k^2}{4\mw^2}\right)^2+8\left(\frac{k^2}{4\mw^2}\right)+6
\right]+\od(\La^{-2})\,,
\eeq

\bea
\De_2(k^2)
&=&
-\frac{i\La_V^4}{96(4\pi)^2\mw^4}
\left[\frac{1}{\ep}+\frac{5}{6}-\ln\frac{\La_V^2}{\mubar^2}\right]
\nn\\
&&
+\frac{i\La_W^2\La_V^2}{48(4\pi)^2(\La_W^2-\La_V^2)\mw^2}
\ln\frac{\La_W^2}{\La_V^2}
-\frac{7i\La_V^2}{1152(4\pi)^2\mw^2}\left(\frac{k^2}{\mw^2}\right)
+\od(\La^0)\,,
\eea

\beq
\mb{Re}\De_2(k^2)
=\frac{\pi\sqrt{1-4\mw^2/k^2}}{60(4\pi)^2}\left[
\left(\frac{k^2}{4\mw^2}\right)-1\right]^2+\od(\La^{-2})\,,
\eeq

\beq
\De_3(k^2)
=\frac{11i\La_V^2}{576(4\pi)^2\mw^2}\left(\frac{k^2}{\mw^2}\right)
+\od(\La^0)\,,
\eeq

\beq
\mb{Re}\De_3(k^2)
=\frac{\pi\sqrt{1-4\mw^2/k^2}}{60(4\pi)^2}\left[
-6\left(\frac{k^2}{4\mw^2}\right)^2-13\left(\frac{k^2}{4\mw^2}\right)+4
\right]+\od(\La^{-2})\,,
\eeq

\beq
\De_4(k^2)
=-\frac{7i\La_V^2}{576(4\pi)^2\mw^2}\left(\frac{k^2}{\mw^2}\right)
+\od(\La^0)\,,
\eeq

\beq
\mb{Re}\De_4(k^2)
=\frac{\pi\sqrt{1-4\mw^2/k^2}}{60(4\pi)^2}
\left[\left(\frac{k^2}{4\mw^2}\right)-1\right]
\left[4\left(\frac{k^2}{4\mw^2}\right)+1\right]
+\od(\La^{-2})\,,
\eeq

\beq
\De_5(k^2)=\od(\La^0)\,,
\eeq

\beq
\label{rede5}
\mb{Re}\De_5(k^2)
=\frac{\pi\sqrt{1-4\mw^2/k^2}}{60(4\pi)^2}\left[
8\left(\frac{k^2}{4\mw^2}\right)^2+4\left(\frac{k^2}{4\mw^2}\right)+3
\right]+\od(\La^{-2})\,,
\eeq
where $\ep$ is defined by $d=4-2\ep$ with spacetime dimension $d$.
Note that $\De_1(k^2)$ and $\De_2(k^2)$ have the same quartically divergent
part
\beq
\De^{(4)}\equiv-\frac{i\La_V^4}{96(4\pi)^2\mw^4}
\left[\frac{1}{\ep}+\frac{5}{6}-\ln\frac{\La_V^2}{\mubar^2}\right]\,.
\eeq
Note further that $\De_3(k^2)$, $\De_4(k^2)$ are only quadratically
divergent and that $\De_5(k^2)$ is at most logarithmically divergent.
In section \ref{transfermatrices} we will need the following linear
combinations which are only quadratically divergent:
\beq
\label{de12}
-4i[\De_1(k^2)-\De_2(k^2)]
=-\frac{\La_V^2}{2(4\pi)^2\mw^2}
\frac{\ln(\La_W^2/\La_V^2)}{1-\La_V^2/\La_W^2}
+\od(\La^0)
\eeq
and
\bea
\label{de1234}
\lefteqn{-4i[\De_1(k^2)-\De_2(k^2)+\De_3(k^2)-\De_4(k^2)]}
\nn\\
&=&
\frac{\La_V^2}{8(4\pi)^2\mw^4}
\left(k^2-4\mw^2\frac{\ln(\La_W^2/\La_V^2)}{1-\La_V^2/\La_W^2}
+\frac{i\pi(k^2+4\mw^2)(k^2-4\mw^2)^{3/2}}{3\sqrt{k^2}\La_V^2}\right)
\nn\\
&&
+\od(0,-2)\,,
\eea
where the shorthand notation
\beq
\label{omn}
\od(m,n)\equiv\mb{Re}\od(\La^m)+\mb{Im}\od(\La^n)
\eeq
has been used.

\subsection{Tensor Structure of $V_{\al\be\ga\de}^{abcd}$
and $P_{\al\be\ga\de}^{abcd}$}
Now let us analyze the tensor structures that can appear in the
bubble sum.
The SU(2) tensors in (\ref{vertex}) and (\ref{ptensor}) are
\beq
\label{su2tensors}
\de_{ab}\de_{cd}\,,\,\de_{ac}\de_{bd}\,,\,\de_{ad}\de_{bc}\,,
\eeq
which can be rewritten into the following linear combinations
corresponding to isospin 0, 1 and 2 contributions $s_0$, $s_1$ and $s_2$,
respectively, in the $(a,b)\lra(c,d)$ channel, where we also
indicate the parity under the symmetry transformations
(\ref{sym1})-(\ref{sym3}).
\bea
\label{s-table}
\ba{lcl|ccc}
&&&a\lra b&c\lra d&(a,b)\lra(c,d)\\
\hline
s_0&=&\ts\frac{1}{3}\de_{ab}\de_{cd}&+&+&+\\
s_1&=&\ts\half(\de_{ac}\de_{bd}-\de_{ad}\de_{bc})&-&-&+\\
s_2&=&\ts-\frac{1}{3}\de_{ab}\de_{cd}
+\half(\de_{ac}\de_{bd}+\de_{ad}\de_{bc})&+&+&+
\ea\,.
\eea
These relations can be inverted to give
\bea
\de_{ab}\de_{cd}&=&3s_0\,,
\\
\de_{ac}\de_{bd}&=&s_0+s_1+s_2\,,
\\
\de_{ad}\de_{bc}&=&s_0-s_1+s_2\,.
\eea
Define the product $s_ks_l$ by $(s_k)_{abmn}(s_l)_{mncd}$.
This can be analyzed in terms of the $s_k$ again such that we have
an algebra
\beq
s_ks_l=\si_{klm}s_m\,.
\eeq
It is easy to see that the $\si_{klm}$ are given by
\beq
\si_{klm}=\left\{\ba{ll}1&k=l=m,\\0&\mb{otherwise,}\ea\right.
\eeq
which merely means that the different isospin channels do not mix.
This can be illustrated by defining the matrix $S$ with elements
$S_{kl}=s_ks_l$,
\beq
S=\left(
\ba{ccc}
s_0&   &   \\
   &s_1&   \\
   &   &s_2
\ea
\right)\,,
\eeq
where empty entries are vanishing, and observing the absence of non-zero
off-diagonal elements.

Going to an isospin basis
\bea
|\pm\ran&\hat{=}&W_\mu^\pm=\ts\frac{1}{\sqrt{2}}(W_\mu^1\mp iW_\mu^2)\,,\\
|0\ran&\hat{=}&Z_\mu=iW_\mu^3\,,
\eea
we can use Clebsch-Gordan coefficients to write
\bea
s_0
&\hat{=}&
|0,0\ran\lan0,0|\,,
\\
s_1
&\hat{=}&
\sum_{k=-1}^1
|1,k\ran\lan1,k|\,,
\\
s_2
&\hat{=}&
\sum_{k=-2}^2|2,k\ran\lan2,k|\,,
\eea
where the first entry means the total isospin and the second its
three-component.

The Lorentz tensors in (\ref{vertex}) and (\ref{ptensor}) are
\beq
\label{lorentz1}
g_{\al\be}g_{\ga\de}\,,\,g_{\al\ga}g_{\be\de}\,,\,g_{\al\de}g_{\be\ga}\,,
\eeq
\beq
\label{lorentz2}
g_{\al\be}k_{\ga}k_{\de}\,,\,g_{\al\ga}k_{\be}k_{\de}
\,,\,g_{\al\de}k_{\be}k_{\ga}\,,\,
g_{\ga\de}k_{\al}k_{\be}\,,\,g_{\be\de}k_{\al}k_{\ga}
\,,\,g_{\be\ga}k_{\al}k_{\de}\,,
\eeq
\beq
\label{lorentz3}
k_{\al}k_{\be}k_{\ga}k_{\de}\,.
\eeq
Define
\beq
g_{\mu\nu}^{tr}=g_{\mu\nu}-k_\mu k_\nu/k^2\,.
\eeq
Then the tensors in (\ref{lorentz1})-(\ref{lorentz3}) can be expressed
in terms of the linear combinations
\bea
t_1
&=&\ts
\frac{1}{d-1}g_{\al\be}^{tr}g_{\ga\de}^{tr}\,,
\\
t_2
&=&\ts
\frac{1}{\sqrt{d-1}k^2}g_{\al\be}^{tr}k_\ga k_\de\,,
\\
t_3
&=&\ts
\frac{1}{\sqrt{d-1}k^2}g_{\ga\de}^{tr}k_\al k_\be\,,
\\
t_4
&=&\ts
\frac{1}{k^4}k_\al k_\be k_\ga k_\de\,,
\\
t_5
&=&\ts
-\frac{1}{d-1}g_{\al\be}^{tr}g_{\ga\de}^{tr}
+\half(g_{\al\ga}^{tr}g_{\be\de}^{tr}+g_{\al\de}^{tr}g_{\be\ga}^{tr})\,,
\\
t_6
&=&\ts
\half(g_{\al\ga}^{tr}g_{\be\de}^{tr}-g_{\al\de}^{tr}g_{\be\ga}^{tr})\,,
\\
t_7
&=&\ts
\frac{1}{2k^2}(g_{\al\ga}^{tr}k_\be k_\de+g_{\be\de}^{tr}k_\al k_\ga
+g_{\al\de}^{tr}k_\be k_\ga+g_{\be\ga}^{tr}k_\al k_\de)\,,
\\
t_8
&=&\ts
\frac{1}{2k^2}(g_{\al\ga}^{tr}k_\be k_\de-g_{\be\de}^{tr}k_\al k_\ga
-g_{\al\de}^{tr}k_\be k_\ga+g_{\be\ga}^{tr}k_\al k_\de)\,,
\\
t_9
&=&\ts
\frac{1}{2k^2}(g_{\al\ga}^{tr}k_\be k_\de-g_{\be\de}^{tr}k_\al k_\ga
+g_{\al\de}^{tr}k_\be k_\ga-g_{\be\ga}^{tr}k_\al k_\de)\,,
\\
t_{10}
&=&\ts
\frac{1}{2k^2}(g_{\al\ga}^{tr}k_\be k_\de+g_{\be\de}^{tr}k_\al k_\ga
-g_{\al\de}^{tr}k_\be k_\ga-g_{\be\ga}^{tr}k_\al k_\de)\,.
\eea
Define the product $t_kt_l$ by
${(t_k)_{\al\be}}^{\mu\nu}(t_l)_{\mu\nu\ga\de}$ and get the algebra
\beq
t_kt_l=\tau_{klm}t_m\,.
\eeq
The only non-zero $\tau_{klm}$ are
\bea
&\tau_{1,1,1}=\tau_{1,2,2}=\tau_{2,3,1}=\tau_{2,4,2}=\tau_{3,1,3}
=\tau_{3,2,4}=\tau_{4,3,3}=\tau_{4,4,4}&
\nn\\
&=\tau_{5,5,5}=\tau_{6,6,6}&
\nn\\
&=\tau_{7,7,7}=\tau_{7,8,8}=\tau_{6,9,7}=\tau_{8,10,8}=\tau_{9,7,9}
=\tau_{9,8,10}=\tau_{10,9,9}=\tau_{10,10,10}=1&.
\eea
To make the structure of this algebra more transparent, define the
matrix $T$ with elements $T_{kl}=t_kt_l$,
\beq
\label{t-table}
T=\left(
\ba{cccccccccc}
\cline{1-4}
\multicolumn{1}{|c}{t_1}&t_2&   &\multicolumn{1}{c|}{}   &&&&&&\\ 
\multicolumn{1}{|c}{}   &   &t_1&\multicolumn{1}{c|}{t_2}&&&&&&\\ 
\multicolumn{1}{|c}{t_3}&t_4&   &\multicolumn{1}{c|}{}   &&&&&&\\ 
\multicolumn{1}{|c}{}   &   &t_3&\multicolumn{1}{c|}{t_4}&&&&&&\\
\cline{1-5}
&&&&\multicolumn{1}{|c|}{t_5}&&&&& \\ 
\cline{5-6}
&&&&&\multicolumn{1}{|c|}{t_6}&&&& \\ 
\cline{6-10}
&&&&&&\multicolumn{1}{|c}{t_7}&  t_8 &   &\multicolumn{1}{c|}{   }   \\ 
&&&&&&\multicolumn{1}{|c}{}   &      &t_7&\multicolumn{1}{c|}{t_8}   \\
&&&&&&\multicolumn{1}{|c}{t_9}&t_{10}&   &\multicolumn{1}{c|}{   }   \\
&&&&&&\multicolumn{1}{|c}{}   &      &t_9&\multicolumn{1}{c|}{t_{10}}\\
\cline{7-10}
\ea
\right)\,.
\eeq
where all empty entries are vanishing and where the lines have been drawn
to guide the eye.

Since the different isospin channels decouple from each other, let us
decompose each relevant tensor $X_{\al\be\ga\de}^{abcd}$ by writing
\beq
X_{\al\be\ga\de}^{abcd}\equiv X=X_0s_0+X_1s_1+X_2s_2\,,
\eeq
e.g.\
\bea
V&=&V_0s_0+V_1s_1+V_2s_2\,,\\
P&=&P_0s_0+P_1s_1+P_2s_2\,,\\
B^{(L)}&=&B_0^{(L)}s_0+B_1^{(L)}s_1+B_2^{(L)}s_2\,.
\eea
The $X_I$ can be decomposed into the $t_k$ by writing
\beq
X_I=X_{Ik}^{(t)}t_k\,,
\eeq
where $k$ runs from 1 through 10.
For example, we have for the anomalous couplings
\bea
V_0
&=&\!\ts
2i\{[(d{+}3)g_4+(3d{-}1)g_5]t_1+\sqrt{d{-}1}(g_4+3g_5)(t_2+t_3)
+5(g_4+g_5)t_4+2(2g_4+g_5)(t_5+t_7)]\}\,,
\nn\\
\\
V_1
&=&\!\ts
-2i\left(g_4-2g_5\right)(t_6+t_{10})\,,
\\
V_2
&=&\!\ts
i\left[2(dg_4+2g_5)t_1+2\sqrt{d-1}g_4(t_2+t_3)+4(g_4+g_5)t_4
+2(g_4+2g_5)(t_5+t_7)\right]
\eea
and for the integrated propagator part
\bea
P_0=P_2
&=&\ts
2(\De_1+d\De_2)t_1
+2\sqrt{d-1}(\De_2+\De_4)(t_2+t_3)
+2(\De_1+2\De_2+2\De_3+4\De_4+\De_5)t_4
\nn\\
&&\ts
+2(\De_1+\De_2)t_5
+2(\De_1+\De_2+\De_3+\De_4)t_7\,,
\\
P_1
&=&\ts
2(\De_1-\De_2)t_6+2(\De_1-\De_2+\De_3-\De_4)t_{10}\,.
\eea
Now we can write 
\beq
\label{bi}
B_I^{(L+1)}=\half\left(B_I^{(L)}P_IV_I+V_IP_IB_I^{(L)}\right)\,.
\eeq
Note that the symmetry (\ref{sym3}) prevents $t_8$ and $t_9$
from appearing in the $V_I$ and $P_I$ and that therefore (\ref{t-table})
tells us that they will not be generated at any point in our calculation.
Equation (\ref{sym3}) also restricts $t_2$ and $t_3$ to appear only in the
combination $t_2+t_3$ in $V_I$, $P_I$ and $B_I^{(L)}$, which is not
the case for the single terms on the right-hand side of (\ref{bi}).

Let us introduce a basis of Lorentz tensors which includes only those
necessary to describe the $V_I$, $P_I$ and $B_I^{(L)}$:
\bea
\label{u-table}
\ba{lcccl|ccc}
&&&&&\al\lra\be&\ga\lra\de&
\left(\!\ba{c}\al\\\be\ea\!\right)\lra\left(\!\ba{c}\ga\\\de\ea\!\right)\\
\hline
u_1&\equiv&t_1&=&\frac{1}{d-1}g_{\al\be}^{tr}g_{\ga\de}^{tr}&+&+&+\\
u_2&\equiv&t_2{+}t_3&=&\frac{1}{\sqrt{d-1}k^2}(g_{\al\be}^{tr}k_\ga k_\de
+g_{\ga\de}^{tr}k_\al k_\be)&+&+&+\\
u_3&\equiv&t_4&=&\frac{1}{k^4}k_\al k_\be k_\ga k_\de&+&+&+\\
u_4&\equiv&t_5&=&-\frac{1}{d-1}g_{\al\be}^{tr}g_{\ga\de}^{tr}
+\half(g_{\al\ga}^{tr}g_{\be\de}^{tr}+g_{\al\de}^{tr}g_{\be\ga}^{tr})
&+&+&+\\
u_5&\equiv&t_6&=&\half(g_{\al\ga}^{tr}g_{\be\de}^{tr}
-g_{\al\de}^{tr}g_{\be\ga}^{tr})&-&-&+\\
u_6&\equiv&t_7&=&
\frac{1}{2k^2}(g_{\al\ga}^{tr}k_\be k_\de{+}g_{\be\de}^{tr}k_\al k_\ga
{+}g_{\al\de}^{tr}k_\be k_\ga{+}g_{\be\ga}^{tr}k_\al k_\de)&+&+&+\\
u_7&\equiv&t_{10}&=&
\frac{1}{2k^2}(g_{\al\ga}^{tr}k_\be k_\de{+}g_{\be\de}^{tr}k_\al k_\ga
{-}g_{\al\de}^{tr}k_\be k_\ga{-}g_{\be\ga}^{tr}k_\al k_\de)&-&-&+
\ea
\eea
As in (\ref{s-table}), we have indicated the parity under the symmetry
transformations (\ref{sym1})-(\ref{sym3}).

To give the tensors $u_1$ through $u_7$ a physical interpretation, let
us consider them in the rest system of $k_\mu$, i.e.\ where
\beq
\bar{k}_\mu\equiv\frac{k_\mu}{\sqrt{k^2}}=(1,0,0,0)
\eeq
and
\beq
-g_{\mu\nu}^{tr}=\left(
\ba{cccc}0&0&0&0\\0&1&0&0\\0&0&1&0\\0&0&0&1\ea
\right)\,.
\eeq
Then the Lorentz index $0$ refers to a spin-0 particle and the other
three components to a spin-1 particle.
In a general Lorentz frame, $W_\mu$ contains a spin-1 field with
$k^\mu W_\mu=0$ and a spin-0 field $k^\mu W_\mu$.
Write the tensors in a bra and ket notation such that
\bea
\kba_\al&\hat{=}&|s\ran_1\\
\kba_\be&\hat{=}&|s\ran_2\\
\kba_\ga&\hat{=}&{}_1\lan s|\\
\kba_\de&\hat{=}&{}_2\lan s|\,,
\eea
(the indices enumerate the particles) where ``s'' refers to a ``scalar,''
and $|1)$, $|2)$, $|3)$ are states with definite spin-up components
in the $x$, $y$ and $z$ direction, respectively.
Going to a spin basis with $|s\ran$ and
\bea
|{\pm}1\ran&=&\ts\frac{1}{\sqrt{2}}[|1)\mp i|2)]\,,\\
|0\ran&=&i|3)\,,
\eea
we can use Clebsch-Gordan coefficients to write
\bea
u_1
&\hat{=}&
|0,0\ran\lan0,0|\,,
\\
u_2
&\hat{=}&
-(|0,0\ran\,{}_1\lan s|{}_2\lan s|+|s\ran_1|s\ran_2\,\lan0,0|)\,,
\\
u_3
&\hat{=}&
|s\ran_1|s\ran_2\,{}_1\lan s|{}_2\lan s|\,,
\\
u_4
&\hat{=}&
\sum_{k=-2}^2|2,k\ran\lan2,k|\,,
\\
u_5
&\hat{=}&
\sum_{k=-1}^1
|1,k\ran\lan1,k|\,,
\\
u_6
&\hat{=}&
-\half\sum_{k=-1}^1
(|k\ran_1|s\ran_2+|s\ran_1|k\ran_2)
({}_1\lan k|{}_2\lan s|+{}_1\lan s|{}_2\lan k|)\,,
\\
u_7
&\hat{=}&
-\half\sum_{k=-1}^1
(|k\ran_1|s\ran_2-|s\ran_1|k\ran_2)
({}_1\lan k|{}_2\lan s|-{}_1\lan s|{}_2\lan k|)\,.
\eea
where the first entry means the total isospin and the second its
three-component, when two entries are present.
This allows for the interpretation of $u_1$, $u_5$ and $u_4$ as channels
for spin-0, 1 and 2 combinations, respectively, from two spin-1 particles.
Then $u_3$ is interpreted as a spin-0 combination of two scalar particles
and $u_2$ as a mixing channel between the spin-0 combination of two spin-1
particles and the spin-0 combination of two scalar particles.
$u_6$ and $u_7$ are different spin-1 combinations of a scalar and a
spin-1 particle.

Now we can write, for the vertex,
\bea
\label{v0u}
V_0
&=&\ts
2i\{[(d+3)g_4+(3d-1)g_5]u_1+\sqrt{d-1}(g_4+3g_5)u_2
+5(g_4+g_5)u_3
+2(2g_4+g_5)(u_4+u_6)]\,,
\nn\\
\\
V_1
&=&\ts
-2i\left(g_4-2g_5\right)(u_5+u_7)\,,
\\
\label{v2u}
V_2
&=&\ts
i\left[2(dg_4+2g_5)u_1+2\sqrt{d-1}g_4u_2+4(g_4+g_5)u_3
+2(g_4+2g_5)(u_4+u_6)\right]
\eea
and, for the integrated propagator part,
\bea
P_0=P_2
&=&\ts
2(\De_1+d\De_2)u_1
+2\sqrt{d-1}(\De_2+\De_4)u_2
+2(\De_1+2\De_2+2\De_3+4\De_4+\De_5)u_3
\nn\\
&&\ts
+2(\De_1+\De_2)u_4
+2(\De_1+\De_2+\De_3+\De_4)u_6\,,
\\
P_1
&=&\ts
2(\De_1-\De_2)u_5+2(\De_1-\De_2+\De_3+\De_4)u_7\,.
\eea

\subsection{Transfer Matrices and Their Eigenvalues}
\label{transfermatrices}
Writing
\beq
B_I^{(L)}=B_{Ik}^{(L)}u_k\,,
\eeq
we can define transfer matrices $M_I$, such that
\beq
B_{Ik}^{(L+1)}=M_{Ikl}B_{Il}^{(L)}\,.
\eeq
Writing
\beq
V_I=V_{Ik}u_k
\eeq
and using that
\beq
B_I^{(0)}=V_I
\eeq
we can write
\beq
\label{symbsum}
B_{Ik}^{(L)}=(M_{I}^L)_{kl}V_{Il}\,.
\eeq
This can be simplified further.
From (\ref{t-table}) and (\ref{u-table}) and the absence of $t_8$ and $t_9$
follows that $u_4$, $u_5$, $u_6$ and $u_7$ propagate independently.
The symmetries (\ref{sym1}) and (\ref{sym2}) together with the parity
properties noted in (\ref{s-table}) and (\ref{u-table}) necessitate
that $u_5$ and $u_7$ appear only in the isospin-1 channel, while the
other $u_k$ appear only in the isospin-0 and isospin-2 channels, which
we summarize in the following table.
\bea
\ba{cc|ccc}
&&\multicolumn{3}{c}{\mb{isospin}}\\
&&0&1&2\\
\hline
&0&s_0u_1,s_0u_2,s_0u_3&&s_2u_1,s_2u_2,s_2u_3\\
\mb{spin}&1&s_0u_6&s_1u_5,s_1u_7&s_2u_6\\
&2&s_0u_4&&s_2u_4
\ea
\eea
Even though $s_1u_5$ and $s_1u_7$ carry the same spin and isospin
assignments, they do not mix.
We can write
\bea
V_0&=&\sum_{k=1}^3V_{0k}^{(123)}u_k+V_0^{(4)}u_4+V_0^{(6)}u_6\,,
\\
V_1&=&V_1^{(5)}u_5+V_1^{(7)}u_7\,,
\\
V_2&=&\sum_{k=1}^3V_{2k}^{(123)}u_k+V_2^{(4)}u_4+V_2^{(6)}u_6\,,
\eea
and then
\bea
B_{0k}^{(L)}
&=&
\sum_{l=1}^3(M_0^{(123)L})_{kl}V_{0l}^{(123)}
\hspace{20pt}k=1,2,3\,,
\\
B_{04}^{(L)}
&=&
\la_0^{(4)L}V_0^{(4)}\,,
\\
B_{06}^{(L)}
&=&
\la_0^{(6)L}V_0^{(6)}\,,
\\
B_{15}^{(L)}
&=&
\la_1^{(5)L}V_1^{(5)}\,,
\\
B_{17}^{(L)}
&=&
\la_1^{(7)L}V_1^{(7)}\,,
\\
B_{2k}^{(L)}
&=&
\sum_{l=1}^3(M_2^{(123)L})_{kl}V_{2l}^{(123)}
\hspace{20pt}k=1,2,3\,,
\\
B_{24}^{(L)}
&=&
\la_2^{(4)L}V_2^{(4)}\,,
\\
B_{26}^{(L)}
&=&
\la_2^{(6)L}V_2^{(6)}\,,
\eea
with
\bea
\la_0^{(4)}
&=&\ts
8i\left(2g_4+g_5\right)(\De_1+\De_2)\,,
\\
\la_0^{(6)}
&=&\ts
8i\left(2g_4+g_5\right)(\De_1+\De_2+\De_3+\De_4)\,,
\eea
\bea
\label{la15}
\la_1^{(5)}
&=&\ts
-4i\left(g_4-2g_5\right)(\De_1-\De_2)\,,
\\
\label{la17}
\la_1^{(7)}
&=&\ts
-4i\left(g_4-2g_5\right)(\De_1-\De_2+\De_3-\De_4)\,,
\eea
\bea
\la_2^{(4)}
&=&\ts
4i\left(g_4+2g_5\right)(\De_1+\De_2)\,,
\\
\la_2^{(6)}
&=&\ts
4i\left(g_4+2g_5\right)(\De_1+\De_2+\De_3+\De_4)\,,
\eea
\beq
M_0^{(123)}=2i
\left(\ba{ccc}
2a_0 &   2c_0  &    0 \\
 b_0 & a_0+d_0 &  c_0 \\
   0 &   2b_0  & 2d_0
\ea\right)\,,
\eeq
\bea
a_0
&=&
[(d+3)g_4+(3d-1)g_5]\De_1+[(d^2+4d-1)g_4+(3d^2+2d-3)g_5]\De_2
\nn\\
&&
+(d-1)(g_4+3g_5)\De_4\,,\\
b_0
&=&
\sqrt{d-1}\{(g_4+3g_5)\De_1+[(d+5)g_4+(3d+5)g_5]\De_2+5(g_4+g_5)\De_4\}\,,
\\
c_0
&=&
\sqrt{d-1}\{(g_4+3g_5)\De_1+[(d+5)g_4+(3d+5)g_5]\De_2
+2(g_4+3g_5)\De_3
\nn\\
&&\hspace{40pt}
+[(d+7)g_4+(3d+11)g_5]\De_4+(g_4+3g_5)\De_5\}\,,
\\
d_0
&=&
5(g_4+g_5)\De_1+[(d+9)g_4+(3d+7)g_5]\De_2+10(g_4+g_5)\De_3
\nn\\
&&
+[(d+19)g_4+(3d+17)g_5]\De_4+5(g_4+g_5)\De_5\,,
\eea
\beq
M_2^{(123)}=2i
\left(\ba{ccc}
2a_2 &   2c_2  &    0 \\
 b_2 & a_2+d_2 &  c_2 \\
   0 &   2b_2  & 2d_2
\ea\right)\,,
\eeq
\bea
a_2
&=&
(dg_4+2g_5)\De_1+[(d^2+d-1)g_4+2dg_5]\De_2+(d-1)g_4\De_4\,,
\\
b_2
&=&
\sqrt{d-1}\{g_4\De_1+[(d+2)g_4+2g_5]\De_2+2(g_4+g_5)\De_4\}\,,
\\
c_2
&=&
\sqrt{d-1}\{g_4\De_1+[(d+2)g_4+2g_5]\De_2+2g_4\De_3
+[(d+4)g_4+2g_5]\De_4+g_4\De_5\}\,,
\\
d_2
&=&
2(g_4+g_5)(\De_1+2\De_3+\De_5)+[(d+3)g_4+4g_5]\De_2
+[(d+7)g_4+8g_5]\De_4\,.
\eea
All other $B_{Ik}^{(L)}$ are zero.

We still have to diagonalize $M_0^{(123)}$ and $M_2^{(123)}$.
Before explicitly doing so, let us finish the formal development.
Assume $M_0^{(123)}$ to have eigenvectors $v_{00}$, $v_{0+}$, $v_{0-}$
with corresponding eigenvalues $\la_{00}$, $\la_{0+}$, $\la_{0-}$.
In the same way, assume $M_2^{(123)}$ to have eigenvectors
$v_{20}$, $v_{2+}$, $v_{2-}$ with corresponding eigenvalues
$\la_{20}$, $\la_{2+}$, $\la_{2-}$.
The $v_{0k}$ and $v_{2k}$ are then linear combinations of the
$u_1$, $u_2$, $u_3$.

If we write the $V_I$ as
\bea
V_0&=&\sum_{k=0,\pm}c_{0k}v_{0k}+c_0^{(4)}u_4+c_0^{(6)}u_6\,,
\\
V_1&=&c_1^{(5)}u_5+c_1^{(7)}u_7\,,
\\
V_2&=&\sum_{k=0,\pm}c_{2k}v_{2k}+c_2^{(4)}u_4+c_2^{(6)}u_6\,,
\eea
we can write
\bea
B_0^{(L)}
&=&
\sum_{k=0,\pm}c_{0k}\la_{0k}^Lv_{0k}
+c_0^{(4)}\la_0^{(4)L}u_4+c_0^{(6)}\la_0^{(6)L}u_6\,,
\\
B_1^{(L)}
&=&
c_1^{(5)}\la_1^{(5)L}u_5+c_1^{(7)}\la_1^{(7)L}u_7\,,
\\
B_2^{(L)}
&=&
\sum_{k=0,\pm}c_{2k}\la_{2k}^Lv_{2k}
+c_2^{(4)}\la_2^{(4)L}u_4+c_2^{(6)}\la_2^{(6)L}u_6\,.
\eea
The final result is then
\beq
\label{sigmaisosum}
\Si=\Si_0s_0+\Si_1s_1+\Si_2s_2
\eeq
with
\bea
\Si_0
&=&
\sum_{L=0}^\infty B_0^{(L)}
=\sum_{k=0,\pm}\frac{c_{0k}}{1-\la_{0k}}v_{0k}
+\frac{c_0^{(4)}}{1-\la_0^{(4)}}u_4+\frac{c_0^{(6)}}{1-\la_0^{(6)}}u_6\,,
\\
\Si_1
&=&
\sum_{L=0}^\infty B_1^{(L)}
=\frac{c_1^{(5)}}{1-\la_1^{(5)}}u_5+\frac{c_1^{(7)}}{1-\la_1^{(7)}}u_7\,,
\\
\label{sigmaiso2}
\Si_2
&=&
\sum_{L=0}^\infty B_2^{(L)}
=\sum_{k=0,\pm}\frac{c_{2k}}{1-\la_{2k}}v_{2k}
+\frac{c_2^{(4)}}{1-\la_2^{(4)}}u_4+\frac{c_2^{(6)}}{1-\la_2^{(6)}}u_6\,.
\eea
A resonance arises if for some $k^2$ an eigenvalue becomes unity. 

From the decomposition of $V_0$, $V_1$, $V_2$ in terms of the $u_k$,
(\ref{v0u})-(\ref{v2u}), we can read off
\bea
c_0^{(4)}
&=&\ts
c_0^{(6)}
=
4i\left(2g_4+g_5\right)\,,
\\
\label{c15c17}
c_1^{(5)}
&=&\ts
c_1^{(7)}
=
-2i\left(g_4-2g_5\right)\,,
\\
c_2^{(4)}
&=&\ts
c_2^{(6)}
=
2i\left(g_4+2g_5\right)\,.
\eea

The eigenvectors of a matrix of the form
\beq
M=
\left(\ba{ccc}
2a & 2c  &  0 \\
 b & a+d &  c \\
 0 & 2b  & 2d
\ea\right)
\eeq
are
\bea
v_0&=&\left(\ba{c}-2c\\a-d\\2b\ea\right)\,,\\
v_\pm&=&\left(
\ba{c}
c\left[a-d\pm\sqrt{(a-d)^2+4bc}\right]\\
2bc\\
b\left[d-a\pm\sqrt{(a-d)^2+4bc}\right]\\
\ea
\right)
\eea
with respective eigenvalues
\bea
\la_0&=&a+d\,,
\\
\la_\pm&=&a+d\pm\sqrt{(a-d)^2+4bc}\,.
\eea
Therefore, the eigenvalues of $M_0^{(123)}$ are
\bea
\la_{00}
&=&\ts
2i(a_0+d_0)\,,
\\
\la_{0\pm}
&=&\ts
2i\left(a_0+d_0\pm\sqrt{(a_0-d_0)^2+4b_0c_0}\right)\,.
\eea
Keeping only quartically divergent terms, they become
\bea
\la_{00}
&=&\ts
2i[(d^2+6d+16)g_4+(3d^2+8d+8)g_5]\De^{(4)}+\od(\La^2)\,,
\\
\la_{0+}
&=&\ts
4i(d+2)[(d+4)g_4+(3d+2)g_5]\De^{(4)}+\od(\La^2)\,,
\\
\la_{0-}
&=&\ts
16i(2g_4+g_5)\De^{(4)}+\od(\La^2)\,.
\eea
In the same way, the eigenvalues of $M_2^{(123)}$ are
\bea
\la_{20}
&=&\ts
2i(a_2+d_2)\,,
\\
\la_{2\pm}
&=&\ts
2i\left(a_2+d_2\pm\sqrt{(a_2-d_2)^2+4b_2c_2}\right)\,.
\eea
Keeping again only quartically divergent terms, they become
\bea
\la_{20}
&=&\ts
2i[(d^2+3d+4)g_4+2(d+4)g_5]\De^{(4)}+\od(\La^2)\,,
\\
\la_{2+}
&=&\ts
4i(d+2)[(d+1)g_4+2g_5]\De^{(4)}+\od(\La^2)\,,
\\
\la_{2-}
&=&\ts
8i(g_4+2g_5)\De^{(4)}+\od(\La^2)\,.
\eea

Now let us consider the rest of the eigenvalues $\la_I^{(k)}$.
Keeping only quartically divergent contributions, we get
\bea
\la_0^{(4)}
&=&\ts
16i\left(g_4+g_5\right)\De^{(4)}+\od(\La^2)\,,
\\
\la_0^{(6)}
&=&\ts
16i\left(g_4+g_5\right)\De^{(4)}+\od(\La^2)\,,
\eea
\bea
\la_2^{(4)}
&=&\ts
8i\left(g_4+2g_5\right)\De^{(4)}+\od(\La^2)\,,
\\
\la_2^{(6)}
&=&\ts
8i\left(g_4+2g_5\right)\De^{(4)}+\od(\La^2)\,.
\eea
In $\la_1^{(5)}$ and $\la_1^{(7)}$ the quartically divergent contributions
cancel.
Combining (\ref{la15}) with (\ref{de12}) and (\ref{la17}) with (\ref{de1234})
and keeping also quadratically divergent contributions and the leading
imaginary part of $\la_1^{(7)}$ gives
\bea
\la_1^{(5)}
&=&
-(g_4-2g_5)\frac{\La_V^2}{2(4\pi)^2\mw^2}
\frac{\ln(\La_W^2/\La_V^2)}{1-\La_V^2/\La_W^2}
+\od(\La^0)\,,
\\
\la_1^{(7)}
&=&
(g_4-2g_5)\frac{\La_V^2}{8(4\pi)^2\mw^4}
\left(k^2-4\mw^2\frac{\ln(\La_W^2/\La_V^2)}{1-\La_V^2/\La_W^2}
+\frac{i\pi(k^2+4\mw^2)(k^2-4\mw^2)^{3/2}}{3\sqrt{k^2}\La_V^2}\right)
\nn\\
&&
+\od(0,-2)\,,
\eea
where we have used again the shorthand notation (\ref{omn}).
Notice that the only eigenvalues that are not quartically divergent are
$\la_1^{(5)}$ and $\la_1^{(7)}$.
In fact, with finite $\La_V$ and $\La_W$ the corresponding integrals
are convergent and dimensional regularization is used merely for
convenience.

\subsection{Resonance}
Combining these results with (\ref{sigmaisosum})-(\ref{sigmaiso2})
as well as (\ref{c15c17}) gives
\bea
\label{sires1}
\Si
&=&
\frac{-2i(g_4-2g_5)s_1u_5}{1+(g_4-2g_5)\frac{\La_V^2}{2(4\pi)^2\mw^2}
\frac{\ln(\La_W^2/\La_V^2)}{1-\La_V^2/\La_W^2}+\od(\La^0)}
\nn\\
&&
+\frac{-2i(g_4-2g_5)s_1u_7}{1-(g_4-2g_5)\frac{\La_V^2}{8(4\pi)^2\mw^4}
\left(k^2-4\mw^2\frac{\ln(\La_W^2/\La_V^2)}{1-\La_V^2/\La_W^2}
+\frac{i\pi(k^2+4\mw^2)(k^2-4\mw^2)^{3/2}}{3\sqrt{k^2}\La_V^2}\right)
+\od(0,-2)}
\nn\\
&&
+\od(\La^{-4})\,.
\eea
Under the assumption that the anomalous couplings dominate the gauge
coupling and the additional assumption that
$|g_4-2g_5|\gg8(4\pi)^2\mw^4/(\La_V^2m_r^2)$
[$m_r$ is the resonance mass; see equation (\ref{mr}) below],
the terms in (\ref{sires1}) are in leading order independent of $g_4$
and $g_5$.
Modeling the second term in (\ref{sires1}) with a Breit-Wigner shape
(and consequently neglecting the $k^2$-dependence of the width term)
gives
\bea
\label{bubblesumfr}
&&
\rule[-30pt]{0pt}{60pt}
\bpi(91,0)(-3,17)
\multiput(33,20)(-4,4){4}{\oval(4,4)[tr]}
\multiput(33,24)(-4,4){4}{\oval(4,4)[bl]}
\multiput(33,20)(-4,-4){4}{\oval(4,4)[br]}
\multiput(33,16)(-4,-4){4}{\oval(4,4)[tl]}
\multiput(53,20)(4,4){4}{\oval(4,4)[tl]}
\multiput(53,24)(4,4){4}{\oval(4,4)[br]}
\multiput(53,20)(4,-4){4}{\oval(4,4)[bl]}
\multiput(53,16)(4,-4){4}{\oval(4,4)[tr]}
\put(15,35){\makebox(0,0)[r]{$a,\al$}}
\put(15,5){\makebox(0,0)[r]{$b,\be$}}
\put(35,20){\circle*{3}}
\put(43,20){\circle{16}}
\put(43,20){\makebox(0,0){$\Sigma$}}
\put(51,20){\circle*{3}}
\put(71,35){\makebox(0,0)[l]{$c,\ga$}}
\put(71,5){\makebox(0,0)[l]{$d,\de$}}
\epi
=-iC_1\left(\frac{\de_{ac}\de_{bd}-\de_{ad}\de_{bc}}{2}\right)
\left(\frac{g_{\al\ga}^{tr}g_{\be\de}^{tr}
-g_{\al\de}^{tr}g_{\be\ga}^{tr}}{2}\right)
\nn\\
&&\hspace{30pt}
+\frac{iC_2}{k^2-m_r^2+im_r\Ga_r}
\left(\frac{\de_{ac}\de_{bd}-\de_{ad}\de_{bc}}{2}\right)
\left(\frac{g_{\al\ga}^{tr}k_\be k_\de+g_{\be\de}^{tr}k_\al k_\ga
-g_{\al\de}^{tr}k_\be k_\ga-g_{\be\ga}^{tr}k_\al k_\de}{2k^2}\right)
+\od(\La^{-4})
\nn\\
\eea
with
\beq
C_1=\frac{4\mw^2}{\La_V^2}
\left(\frac{\ln(\La_W^2/\La_V^2)}{1-\La_V^2/\La_W^2}\right)^{-1}\,,
\eeq
\beq
C_2=\frac{16(4\pi)^2\mw^4}{\La_V^2}\,,
\eeq
\beq
\label{mr}
m_r=2\mw\sqrt{\frac{\ln(\La_W^2/\La_V^2)}{1-\La_V^2/\La_W^2}}
+\od(\La^{-2})\,,
\eeq
and
\beq
\label{Gammar}
\Ga_r=\frac{\pi(m_r^2+4\mw^2)(m_r^2-4\mw^2)^{3/2}}{3m_r^2\La_V^2}
+\od(\La^{-4})\,.
\eeq
Note that we can trivially drop the superscript ``$tr$'' in the $C_2$
term in (\ref{bubblesumfr}).

The direct calculation of the resonant part of the bubble sum
(\ref{bubblesum}) gives the mass (\ref{mr}) and the width (\ref{Gammar})
for the resonance.
As a consistency check for the width, we have computed the decay rate of
the resonance independently.
For $k^2\approx m_r^2$, the resonant part of the bubble sum
(\ref{bubblesumfr}) can be written as
\beq
\la\ep_{abm}[k_\al g_{\be\mu}-k_\be g_{\al\mu}]
(-i)\de_{mn}\left(\frac{g^{\mu\nu}}{k^2-m_r^2+im_r\Ga_r}+Xk^\mu k^\nu\right)
\la\ep_{cdn}[(-k)_\ga g_{\de\nu}-(-k)_\de g_{\ga\nu}]
\eeq
with
\beq
\la=\frac{2(4\pi)\mw^2}{\La_Vm_r}
\eeq
and arbitrary $X$.
That is, the bubble sum can be decomposed into a spin-1 propagator part
\beq
-i\de_{mn}\left(\frac{g_{\mu\nu}}{k^2-m_r^2+im_r\Ga_r}+Xk_\mu k_\nu\right)
\eeq
and two derivative couplings
\beq
\la\ep_{abc}(k_\al g_{\be\ga}-k_\be g_{\al\ga})\,.
\eeq
This coupling can be used to compute the decay rate into two $W$ bosons.
The result of this standard calculation coincides with (\ref{Gammar}).

\section{Discussion}
\label{discussion}
The first thing one notices about the contributions to the bubble sum is
that in most channels quartic divergences are present.
In these cases we did not calculate the subleading terms.
In case quartic divergences are present the subleading terms could be of the
form $\La^2k^2$.
If such terms are absent the whole interaction is suppressed by $1/\La^4$
and the channel is of no phenomenological interest.
If such terms are present resonances are present at a scale $\od(\La^2)$.
These are out of the reach of present colliders and possibly also of the LHC.
Much more interesting is the $I=1$, $J=1$ channel where we find a low-lying
resonance in the term $u_7$, which corresponds to longitudinal vector boson
scattering.
Depending on the ratio $\La_W/\La_V$ one finds the resonance below
or above the vector boson threshold.
On physical grounds we expect $\La_V$ to be the smaller, as it is
directly related to the Goldstone boson sector of the theory, where the strong
interactions are supposed to take place.
In that case the resonance always lies above the two-$W$-threshold.
Because the interactions of the transversal vector bosons are suppressed by
the gauge coupling, a reasonable assumption would be
$\La_V\approx g\La_W$.
This corresponds to $m_r\approx200\mb{GeV}$.
A recent comparison with the LEP-100 data gives a limit
$\La_V>490\mb{GeV}$ \cite{kast}.
For $m_r=200\mb{GeV}$, this gives $\Ga_r<12 \mb{GeV}$.
The fact that the resonance can be at such low energy is somewhat surprising,
given our experience with chiral perturbation theory in pion physics.

To study the connection with pion physics we make the substitutions
$g_4=g^4\ep_4$, $g_5=g^4\ep_5$ and $\mw=gf_{\pi}/2$.
In the resulting Lagrangian one takes $g\rightarrow 0$, with
$\ep_4$, $\ep_5 $  and $f_{\pi}$ fixed.
This way one finds the standard nonlinear sigma model with two higher
derivative terms.
We define $g_{\rho}=4(\ep_4-2\ep_5)$.
For didactical purposes we keep in the chiral perturbation theory here the
tree level terms, the imaginary part of the ordinary chiral loop and the
contribution of the first loop we calculated.
One finds for the $I=1$, $J=1$ amplitude,
\beq
a_{11}(s)
=\frac{s}{96\pi f_{\pi}^2}\left[1+\frac{g_{\rho}s}{f_{\pi}^2}
+\frac{g_{\rho}^2\La_V^2}{32\pi^2f_{\pi}^2}\frac{s^2}{f_{\pi}^4}
+\frac{is}{96\pi f_{\pi}^2}\left(1+\frac{g_{\rho}^2s^2}{f_{\pi}^4}
\right)\right]\,.
\eeq
After unitarizing the amplitude with the $[1,1]$ Pad\'e approximant
one finds the $\rho$ resonance with a width given by
\beq
\Gamma_{\rho}
=\frac{m_{\rho}^3}{96\pi f_{\pi}^2}
\left(1-\frac{\La_V^2}{32\pi^2f_{\pi}^2}\right)\,.
\eeq
We had to ignore the $s^4$ term in the imaginary part of $a_{11}$
because it is of too high an order in chiral perturbation theory and other
terms of this order have been ignored.
The presence of the extra loops that we calculated results therefore
in a correction to the KSRF \cite{ksrf} relation
$\Ga_{\rho}=m_{\rho}^3/(96\pi f_{\pi}^2)$.
Of course, this is not the full story for chiral perturbation theory, but it
shows that the effects we calculated are only a small correction to pion
physics.
The reason we get a large effect in our calculation is that we assume that
the interactions are dominated by the $g_{\rho}$ interaction.
In the chiral limit this is not possible, because at low enough energy the
lowest order term in $s$ always dominates.
Since within the standard model there is the $W$ threshold to consider, it
makes sense to say that the anomalous term dominates.
It means $|g_{\rho}|s^2/v^2\gg s$ for $s>4\mw^2$.
Within the standard model this translates into $4|g_4-2g_5|\gg g^2$.
This condition therefore quantifies what we mean by strong anomalous
couplings.

The model we discussed fits in naturally in the class of models that give
rise to resonances in $W$ physics due to strong interactions.
This class of models is collectively known as the BESS model, from breaking
electroweak symmetry strongly.
A recent review of this class of models is \cite{dominici}.
Nonetheless, as the discussion above points out, the model is subtly
different from the models in the literature, due to the dominance of the
anomalous couplings.
This would make the model into a mere curiosity were it not for  a very
simple class of models giving rise to precisely the required structure.
This is the class of the strongly interacting singlet Higgs (SISH) models
\cite{hillvelt}.
These models, which contain beyond the standard model only extra scalar
singlet particles, are actually the  simplest possible renormalizable
extensions of the standard model.
Because the singlets couple only to the Higgs boson, they do not change
the phenomenology at LEP-100 at the one-loop level.
Two-loop effects are too small to  be significant.
In order to perform a precise phenomenology of the model, SU$_R$(2) breaking
effects should be taken into account.
The model satisfies the LEP-100 limits on extra $Z$ bosons, because both
the coupling to leptons and the mixing with the $Z$ boson are suppressed
by an electroweak loop.
The statistics at the Tevatron is too small to see the resonance. 
For the planned high-energy colliders the phenomenology should be
straightforward, the resonances being produced via vector boson fusion.
For LEP-200 the situation is somewhat subtle, as the resonance does not
couple to the incoming electrons directly.
Strong form factor effects could play a role.
The precise phenomenology will be left to future work.

Given the fact that the couplings are strong, one can question how generic
the results are.
In principle, higher-order terms in the chiral perturbation theory could
be important.
For the formation of the resonance via a bubble sum, only four-point
vertices contribute.
The presence of higher order terms will therefore not effect the structure
of the calculation very much.
Basically, we expect a form factor for $g_4-2g_5$ which, when inserted
into the graphs, makes the explicit dependence on $\La_V$ more
complicated.
However, the term $(g_4-2g_5)\Lambda_V^2$ in formula (\ref{sires1}) is
essentially determined by power counting; so one would expect corrections
of the form $\La^2\rightarrow\La^2+\od(4 m_W^2)$.
Also, the precise formula for the resonance mass as a function of
$\La_W$, $\La_V$ could become more complicated.
However that will not change the qualitative picture of
$m_r\approx 200 GeV$, with a coupling of order  $m_r/\La_V$.

Finally one can ask if the model is consistent with the LEP precision data.
The limits derived on $g_4$, $g_5$ in \cite{quart,kast} follow from simple
perturbation theory and depend crucially on the behavior in the
hypercharge sector of the theory, as they come from the limits on the
$\rho$ parameter.
For the strongly interacting case as described here they are not expected
to be a good estimate and could possibly even be an order of magnitude wrong. 
The model is similar in appearance to the standard model in the large Higgs
boson mass limit, which is disfavored by the LEP1 data.
When one simply removes the Higgs boson from the standard model, the model
becomes nonrenormalizable, but the radiative effects grow only
logarithmically with the cutoff.
The question is whether this scenario is ruled out by the LEP1 precision
data.
The LEP1 data appear to be in agreement with the standard model, with a
preferred low Higgs boson mass.
One is sensitive to the Higgs boson mass in three parameters, known as
$S$, $T$, $U$ or $\ep_1$, $\ep_2$, $\ep_3$.
They receive corrections of the form $g^2[\ln(m_H/m_W)+\mb{const}]$, where
the constants are of order 1.
The logarithmic enhancement is universal and would also appear in models
without a Higgs boson as $\ln(\La)$, where $\La$ is the cutoff where new
interactions should appear.
Only when one can determine the three different constants can one say
one has established the standard model.
At present, the data do not suffice to do this to sufficiently high precision.
In practice, one can compensate a change in the Higgs boson mass in the
formulas with extra contributions to the $S$, $T$ and $U$ parameters.
As such terms are generated by the contributions of formula (\ref{hicoder}),
there is enough freedom to fit the data; see \cite{kast} for a discussion.
Whether a model with a low-lying resonance would actually improve the fit 
to the data depends on the couplings to the fermions and to the hypercharge.

\section*{Acknowledgements}
We thank G.~Jikia and T.N.~Truong for discussions on chiral perturbation
theory.
B.K.\ also would like to thank R.~Akhoury, T.~Binoth, M.~Chanowitz, T.~Clark,
A.~Denner, D.~Graudenz, S.~Khlebnikov, S.~Love, B.~Tausk and D.~Wackeroth
for useful discussions.
This work was supported by the Deutsche Forschungsgemeinschaft (DFG).

\appendix

\section{One-Loop Results}
\label{oneloop}

Here we list results for the necessary basic one-loop integrals and for
the composite one-loop quantities $\De_1$ through $\De_5$.

Let $d$ be the spacetime dimension, $d=4-2\ep$, let $\mubar$ be given by
$\ln4\pi\mu^2-\ga_{\sss E}=\ln\mubar^2$ with the Euler-Mascheroni constant
$\ga_{\sss E}$  and use the abbreviation $\int_p$ given in (\ref{int})\,.

The only integrals we need are
\beq
I(m^2)\equiv\int_p\frac{1}{p^2-m^2}=
\frac{im^2}{(4\pi)^2}\left(\frac{1}{\ep}+1-\ln\frac{m^2}{\mubar^2}\right)
+\od(\ep)
\eeq
and 
\bea
\label{ikmambdef}
I(k^2;m_a^2,m_b^2)
&\equiv&
\int_p\frac{1}{[(p+k)^2-m_a^2+i\ve](p^2-m_b^2+i\ve)}\,.
\eea
Define
\beq
\label{D}
D\equiv k^4+m_a^4+m_b^4-2k^2m_a^2-2k^2m_b^2-2m_a^2m_b^2\,.
\eeq
For $k^2\geq0$ we get the following results in a straightforward calculation.

\subsection{$I(k^2;m_a^2,m_b^2)$ for $D\leq0$}
$D\leq 0$ is equivalent to the statement that none of $k^2$, $m_a^2$, $m_b^2$
is larger than the sum of the other two.
\bea
\label{ikmm1}
I(k^2;m_a^2,m_b^2)
&=&
\frac{i}{(4\pi)^2}\Bigg[\frac{1}{\ep}+2
-\frac{k^2+m_a^2-m_b^2}{2k^2}\ln\frac{m_a^2}{\mubar^2}
-\frac{k^2+m_b^2-m_a^2}{2k^2}\ln\frac{m_b^2}{\mubar^2}
\nn\\
&&\hspace{35pt}
-\frac{\sqrt{-D}}{k^2}\left(\arctan\frac{k^2+m_a^2-m_b^2}{\sqrt{-D}}
+\arctan\frac{k^2+m_b^2-m_a^2}{\sqrt{-D}}\right)
\Bigg]+\od(\ep)\,.
\eea

\subsection{$I(k^2;m_a^2,m_b^2)$ for $D\geq0$ with $k^2\leq|m_a^2-m_b^2|$}
\bea
\label{ikmm2}
\lefteqn{I(k^2;m_a^2,m_b^2)}
\nn\\
&=&
\frac{i}{(4\pi)^2}\left\{\frac{1}{\ep}+2
+\frac{m_b^2-m_a^2-k^2}{2k^2}\ln\frac{m_a^2}{\mubar^2}
+\frac{m_a^2-m_b^2-k^2}{2k^2}\ln\frac{m_b^2}{\mubar^2}
+\frac{\sqrt{D}}{k^2}
\ln\frac{m_a^2+m_b^2-k^2+\sqrt{D}}{m_a^2+m_b^2-k^2-\sqrt{D}}
\right\}
\nn\\
&&
+\od(\ep)\,.
\eea
Equation (\ref{ikmm2}) can also be written as
\bea
\label{ikmm2a}
I(k^2;m_a^2,m_b^2)
&=&
\frac{i}{(4\pi)^2}\Bigg\{\frac{1}{\ep}+2
+\frac{m_b^2-m_a^2-k^2+\sqrt{D}}{2k^2}\ln\frac{m_a^2}{\mubar^2}
+\frac{m_a^2-m_b^2-k^2+\sqrt{D}}{2k^2}\ln\frac{m_b^2}{\mubar^2}
\nn\\
&&\hspace{35pt}
+\frac{\sqrt{D}}{k^2}
\ln\frac{m_a^2+m_b^2-k^2+\sqrt{D}}{2\mubar^2}
\Bigg\}+\od(\ep)
\eea
or
\bea
\label{ikmm2b}
I(k^2;m_a^2,m_b^2)
&=&
\frac{i}{(4\pi)^2}\Bigg\{\frac{1}{\ep}+2
+\frac{m_b^2-m_a^2-k^2-\sqrt{D}}{2k^2}\ln\frac{m_a^2}{\mubar^2}
+\frac{m_a^2-m_b^2-k^2-\sqrt{D}}{2k^2}\ln\frac{m_b^2}{\mubar^2}
\nn\\
&&\hspace{35pt}
-\frac{\sqrt{D}}{k^2}
\ln\frac{m_a^2+m_b^2-k^2-\sqrt{D}}{2\mubar^2}
\Bigg\}+\od(\ep)\,.
\eea

\subsection{$I(k^2;m_a^2,m_b^2)$ for $D\geq0$ with $k^2\geq |m_a^2-m_b^2|$}
Here
\bea
\label{ikmm3}
\lefteqn{I(k^2;m_a^2,m_b^2)}
\nn\\
&=&
\frac{i}{(4\pi)^2}\Bigg\{\frac{1}{\ep}+2
+\frac{m_b^2-m_a^2-k^2}{2k^2}\ln\frac{m_a^2}{\mubar^2}
+\frac{m_a^2-m_b^2-k^2}{2k^2}\ln\frac{m_b^2}{\mubar^2}
+\frac{\sqrt{D}}{k^2}
\ln\frac{k^2-m_a^2-m_b^2-\sqrt{D}}{k^2-m_a^2-m_b^2+\sqrt{D}}
\Bigg\}
\nn\\
&&
-\frac{\sqrt{D}}{16\pi k^2}+\od(\ep)\,.
\eea

\subsection{$I(k^2;m^2,m^2)$}
If $D=k^2(k^2-4m^2)\leq0$, i.e., for $k^2\leq4m^2$, we get from (\ref{ikmm1})
\beq
\label{ikmama}
I(k^2;m^2,m^2)=
\frac{i}{(4\pi)^2}\left(\frac{1}{\ep}+2-\ln\frac{m^2}{\mubar^2}
-2{\ts\sqrt{\frac{4m^2}{k^2}-1}}\;
\arctan\frac{1}{\sqrt{\frac{4m^2}{k^2}-1}}\right)+\od(\ep)\,.
\eeq
For $k^2\ll m^2$, we can expand in powers of $k^2/m^2$ to get
\beq
I(k^2;m^2,m^2)=
\frac{i}{(4\pi)^2}\left[\frac{1}{\ep}-\ln\frac{m^2}{\mubar^2}
+\frac{1}{6}\left(\frac{k^2}{m^2}\right)
+\frac{1}{60}\left(\frac{k^2}{m^2}\right)^2\right]
+\od\left(\ep,\left(\ts\frac{k^2}{m^2}\right)^3\right)\,.
\eeq

If $D=k^2(k^2-4m^2)\geq0$, i.e., for $k^2\geq4m^2$, (\ref{ikmm3}) becomes
\beq
I(k^2;m^2,m^2)=
\frac{i}{(4\pi)^2}\left[
\frac{1}{\ep}+2-\ln\frac{m^2}{\mubar^2}
+{\ts\sqrt{1-\frac{4m^2}{k^2}}}
\ln\frac{1-\frac{2m^2}{k^2}-\sqrt{1-\frac{4m^2}{k^2}}}
{1-\frac{2m^2}{k^2}+\sqrt{1-\frac{4m^2}{k^2}}}
\right]
-\frac{\sqrt{1-\frac{4m^2}{k^2}}}{16\pi}+\od(\ep)\,.
\eeq

\subsection{$I(k^2;m^2,0)$ for $k^2\leq m^2$}
Now $D=|k^2-m^2|\geq0$.
Let us assume $k^2\leq m^2$.
Then we get from (\ref{ikmm2a}) or (\ref{ikmm2b})
\bea
\label{ikm0}
I(k^2;0,m^2)=I(k^2;m^2,0)
&=&
\frac{i}{(4\pi)^2}\left[\frac{1}{\ep}+2
+\frac{m^2-k^2}{k^2}\ln\frac{m^2-k^2}{\mubar^2} 
-\frac{m^2}{k^2}\ln\frac{m^2}{\mubar^2}\right]+\od(\ep)
\nn\\
&=&
\frac{i}{(4\pi)^2}\left[\frac{1}{\ep}+2
+\frac{m^2}{k^2}\ln\left(1-\frac{k^2}{m^2}\right) 
-\ln\frac{m^2-k^2}{\mubar^2}\right]+\od(\ep)\,.
\eea
For $k^2\ll m^2$, we can expand in powers of $k^2/m^2$ to get
\bea
\lefteqn{I(k^2;0,m^2)=I(k^2;m^2,0)}
\nn\\
&=&
\frac{i}{(4\pi)^2}\left[\frac{1}{\ep}+1-\ln\frac{m^2}{\mubar^2}
+\frac{1}{2}\left(\frac{k^2}{m^2}\right)
+\frac{1}{6}\left(\frac{k^2}{m^2}\right)^2
+\frac{1}{12}\left(\frac{k^2}{m^2}\right)^3
+\frac{1}{20}\left(\frac{k^2}{m^2}\right)^4
\right]+\od(\ep,\left(\ts\frac{k^2}{m^2}\right)^5)\,.
\nn\\
\eea

\subsection{$I(k^2;m_a^2,m_b^2)$ for $k^2,m_a^2\ll m_b^2$}
If $k^2,m_a^2\ll m_b^2$, we can expand (\ref{ikmm2a}) or (\ref{ikmm2b})
in negative powers of $m_b^2$ to get
\bea
\lefteqn{I(k^2;m_a^2,m_b^2)}
\nn\\
&=&
\frac{i}{(4\pi)^2}
\bigg[\frac{1}{\ep}+1-\ln\frac{m_b^2}{\mubar^2}
+\frac{\frac{1}{2}k^2+m_a^2\ln\frac{m_a^2}{m_b^2}}{m_b^2}
+\frac{k^2(\frac{1}{6}k^2+\frac{3}{2}m_a^2)+
m_a^2(k^2+m_a^2)\ln\frac{m_a^2}{m_b^2}}{m_b^4}
\nn\\
&&\hspace{35pt}
+\frac{k^2(\frac{1}{12}k^4+\frac{7}{3}k^2m_a^2+\frac{5}{2}m_a^4)
+m_a^2(k^4+3k^2m_a^2+m_a^4)\ln\frac{m_a^2}{m_b^2}}{m_b^6}
\nn\\
&&\hspace{35pt}
+\frac{k^2(\frac{1}{20}k^6+\frac{35}{12}k^4m_a^2
+\frac{17}{2}k^2m_a^4+\frac{7}{2}m_a^6)
+m_a^2(k^6+6k^4m_a^2+6k^2m_a^4+m_a^6)\ln\frac{m_a^2}{m_b^2}}{m_b^8}
\bigg]
\nn\\
&&
+\od(\ep,m_b^{-10}\ln m_b^2)\,.
\eea

\subsection{$I(k^2;m_a^2,m_b^2)$ for $k^2\ll m_a^2,m_b^2$}
Here
\bea
\label{ik2mambex}
\lefteqn{I(k^2;m_a^2,m_b^2)}
\nn\\
&=&
\frac{i}{(4\pi)^2}\Bigg[\frac{1}{\ep}+1
-\frac{m_a^2\ln\frac{m_a^2}{\mubar^2}
-m_b^2\ln\frac{m_b^2}{\mubar^2}}{m_a^2-m_b^2}
+\Bigg(\frac{m_a^2+m_b^2}{2(m_a^2-m_b^2)^2}
-\frac{m_a^2m_b^2}{(m_a^2-m_b^2)^3}\ln\frac{m_a^2}{m_b^2}\Bigg)k^2
\nn\\
&&\hspace{35pt}
+\Bigg(\frac{m_a^4+10m_a^2m_b^2+m_b^4}{6(m_a^2-m_b^2)^4}
-\frac{m_a^2m_b^2(m_a^2+m_b^2)}{(m_a^2-m_b^2)^5}
\ln\frac{m_a^2}{m_b^2}\Bigg)k^4
\nn\\
&&\hspace{35pt}
+\Bigg(\frac{m_a^6+29m_a^4m_b^2+29m_a^2m_b^4+m_b^6}{12(m_a^2-m_b^2)^6}
-\frac{m_a^2m_b^2(m_a^4+3m_a^2m_b^2+m_b^4)}{(m_a^2-m_b^2)^7}
\ln\frac{m_a^2}{m_b^2}\Bigg)k^6
\Bigg]
\nn\\
&&
+\od(k^8,\ep)\,.
\eea

\subsection{Calculation of $\De_1$, $\De_2$, $\De_3$, $\De_4$, $\De_5$}
To compute the quantities $\De_1$ through $\De_5$ defined
in (\ref{ptensor}), we first need some preliminaries.
Let $f$ in the following be some function of $p^2$ and $(p+k)^2$.

Contracting
\beq
\Int{p}p_\al f=A\frac{k_\al}{k^2}
\eeq
with $k_\al$ gives
\beq
A=\Int{p}(p\cdot k)f\,.
\eeq

Contracting
\beq
\Int{p}p_\al p_\be f=A'g_{\al\be}+B'\frac{k_\al k_\be}{k^2}
\eeq
with $g_{\al\be}$ and $k_\al k_\be$ and solving the resulting equations
for $A'$ and $B'$ we get
\bea
A'
&=&\ts
\frac{1}{d-1}\Int{p}\left(p^2-\frac{(p\cdot k)^2}{k^2}\right)f\,,
\\
B'
&=&\ts
\frac{1}{d-1}\Int{p}\left(-p^2+d\frac{(p\cdot k)^2}{k^2}\right)f\,.
\eea

Contracting
\beq
\Int{p}p_\al p_\be p_\ga f
=A''\frac{k_\al g_{\be\ga}+k_\be g_{\ga\al}+k_\ga g_{\al\be}}{k^2}
+B''\frac{k_\al k_\be k_\ga}{k^4}
\eeq
with $k_\al g_{\be\ga}$ and $k_\al k_\be k_\ga$ and solving the resulting
equations for $A''$ and $B''$, we get
\bea
A''
&=&\ts
\frac{1}{d-1}\Int{p}\left(p^2(p\cdot k)-\frac{(p\cdot k)^3}{k^2}\right)f\,,
\\
B''
&=&\ts
\frac{1}{d-1}\Int{p}\left(-3p^2(p\cdot k)+(d+2)\frac{(p\cdot k)^3}{k^2}
\right)f\,.
\eea

Contracting
\bea
\Int{p}p_\al p_\be p_\ga p_\de f
&=&\ts
A'''(g_{\al\be}g_{\ga\de}+g_{\al\ga}g_{\be\de}+g_{\al\de}g_{\be\ga})
+B'''
\frac{\ba{l}\scs g_{\al\be}k_\ga k_\de+g_{\al\ga}k_\be k_\de
+g_{\al\de}k_\be k_\ga
\vspace{-4pt}\\
\scs\;+g_{\ga\de}k_\al k_\be+g_{\be\de}k_\al k_\ga
+g_{\be\ga}k_\al k_\de\ea}{k^2}
+C'''\frac{k_\al k_\be k_\ga k_\de}{k^4}
\nn\\
\eea
with $g_{\al\be}g_{\ga\de}$, $k_\al k_\be g_{\ga\de}$ and
$k_\al k_\be k_\ga k_\de$ and solving the resulting equations for $A'''$,
$B'''$ and $C'''$ we get
\bea
A'''
&=&\ts
\frac{1}{(d-1)(d+1)}
\Int{p}\left(p^4-2\frac{p^2(p\cdot k)^2}{k^2}+\frac{(p\cdot k)^4}{k^4}
\right)f\,,
\\
B'''
&=&\ts
\frac{1}{(d-1)(d+1)}
\Int{p}\left(-p^4+(d+3)\frac{p^2(p\cdot k)^2}{k^2}
-(d+2)\frac{(p\cdot k)^4}{k^4}\right)f\,,
\\
C'''
&=&\ts
\frac{1}{(d-1)(d+1)}
\Int{p}\left(3p^4-6(d+2)\frac{p^2(p\cdot k)^2}{k^2}
+(d+2)(d+4)\frac{(p\cdot k)^4}{k^4}\right)f\,.
\eea

Using the above results, some trivial but lengthy algebra on the integrand,
the properties of dimensional regularization, the results for one-loop
integrals in the preceding parts of the appendix, as well as
\beq
\int_p(p\cdot k)^{2n}f(p^2)=c_n(k^2)^n\int_p(p^2)^n f(p^2)
\eeq
with
\beq
c_0=1\,,\;\;\;\;\;\;c_n=\frac{2n-1}{d+2n-2}c_{n-1}\,,
\eeq
i.e.\
\beq
c_n=\frac{\Ga(n+1/2)\Ga(d/2)}{\Ga(n+d/2)\Ga(1/2)}\,,
\eeq
we can compute $\De_1$ through $\De_5$.
The results are given in the main text in eqs.\ (\ref{de1}) through
(\ref{rede5}).

\end{document}